# Au-Ag template stripped pattern for scanning probe investigations of DNA arrays produced by Dip Pen Nanolithography

*Andrea Baserga[1,2], Marco Vigano'[1], Carlo S. Casari[1,2]\*, Stefano Turri[1], Andrea Li Bassi[1,2], Marinella Levi[1], Carlo E. Bottani[1,2]*

[1]Dipartimento di Chimica, Materiali e Ingegneria Chimica 'G. Natta', Politecnico di Milano, Piazza Leonardo da Vinci 32, I-20133 Milano, Italy

[2]NEMAS-Center for NanoEngineered Materials and Surfaces and IIT-Italian Institute of Technology, Politecnico di Milano, Italy

\* Corresponding author. E-mail: carlo.casari@polimi.it

ABSTRACT

We report on DNA arrays produced by Dip Pen Nanolithography (DPN) on a novel Au-Ag micro patterned template stripped surface. DNA arrays have been investigated by atomic force microscopy (AFM) and scanning tunnelling microscopy (STM) showing that the patterned template stripped substrate enables easy retrieval of the DPN-functionalized zone with a standard optical microscope permitting a multi-instrument and multi-technique local detection and analysis. Moreover the smooth surface of the Au squares (~ 5-10 Å roughness) allows to be sensitive to the hybridization of the oligonucleotide array with label-free target DNA. Our Au-Ag substrates, combining the retrieving capabilities of the patterned surface with the smoothness of the template stripped technique, are candidates for the investigation of DPN nanostructures and for the development of label free detection methods for DNA nanoarrays based on the use of scanning probes.

INTRODUCTION

The increased need for gene expression analysis[1-3] and genotyping[4,5] is driving the biotechnology towards nanotechnology. Atomic force and scanning tunneling microscopy (AFM/STM) allow direct manipulation and investigation of biomolecules (e.g. DNA and proteins) with unprecedented spatial resolution.[6-10] The advantage of biosensors, microarrays and gene chips miniaturization is the possibility to increase the probe material density by several orders of magnitude,[11,12] opening the way for screening molecular targets or even the whole human genome for single-nucleotide polymorphisms with reduced sample volumes and time analysis.[13] In this framework the detection of label-free DNA in nanoarrays by means of scanning probe techniques is of great interest for the development of fast and cost-effective analysis.[14,15]

Dip-Pen nanolithography (DPN) is an AFM-derived technique recently emerged as a tool for the fabrication of multi-component nanostructures on a flat surface with near perfect registration capabilities.[11,16] In this technique an AFM tip, coated with the molecules of interest ("ink"), releases the ink onto the substrate, thus permitting the writing of molecular patterns at a sub-micrometer scale.[16-18] DPN offers the advantage of an extreme diversification of geometrical structures and immobilization chemistries; it works in ambient conditions with the possibility to parallelize the writing process.[19] These features make DPN extremely attractive for the realization of DNA micro and nanoarrays even though some critical points still need to be handled. So far, in most works about DPN-based micro- and nanoarrays the same instrument or even the same tip was used for both writing and imaging. The need to retrieve the microscopically patterned zone (i.e. localization of a patterned region of less than 10 $\mu m^2$ over a sample area of at least $10^3$ $\mu m^2$) once the tip has been retracted and the sample removed from the instrument is a crucial point if one wants to perform different investigations of the deposited biomolecule pattern (e.g. fluorescence, IR absorption, Raman, mass spectrometry) or at least local or selective chemical treatments and functionalizations. This is even more critical if one wants to develop investigation or detection and sensing methods for biomolecule arrays with a real nanometer resolution, e.g. by using local scanning probe microscopy or spectroscopy techniques (AFM/STM, Scanning tunneling spectroscopy,[20] Kelvin probe microscopy,[14] SNOM[21] or near-field Raman microscopy,[22] etc.). This is of crucial importance both for fundamental proteomics/genomics studies, and because it is in principle possible to imagine parallel, high speed multi-tip detection modes based on scanning probes.[23]

Another key issue for the production of DNA micro and nanoarrays is related with the capability to detect DNA hybridization, without need for 'modifying' the target molecule, i.e. by labeling or marking. Label free detection methods are desirable because they simplify the processing and detect the molecule binding in a more native-like state.[24] The actual resolution and sensitivity in detecting DNA hybridization is usually limited by the substrate surface roughness which can be of the same order or higher than the imaged molecule. Some recent works report on the fabrication of DNA nanoarrays,[25,26] but in all cases a chemical modification of the probe DNA (or oligonucleotide) was needed in order to recognize the occurred hybridization, for example gold nanospheres linked to the DNA terminus. Only recently Kelvin probe force microscopy has been proposed for high-resolution, label-free analysis of protein and DNA nanoarrays.[14] On the other hand traditional detection methods (e.g. fluorescence) are based on the use of markers. Strategies aimed at overcoming the need to functionalize the target DNA are of great technological interest for the development of DNA nanoarray chips.

In this work we address the above mentioned issues by means of a multi-metal patterned microgrid substrate based on the template stripping (TS) technique.[27] By combining TS technique with metal evaporation through a grid mask we propose an easy way to produce a substrate with a twofold purpose: on one side to have an Au-Ag micro patterned grid with Au squares visible by a standard optical microscope and on the other to have a highly smoothed gold surface. We here demonstrate that our multi-metal patterned surface easily allows DPN writing and retrieval and detection of the DPN feature of interest, that can thus be investigated by means of different scanning probe techniques. In particular we report on the investigation of DPN-generated DNA arrays with different instruments and different techniques such as contact mode and phase contrast AFM and STM. Furthermore the high surface smoothness achieved by the TS technique permits to be sensitive to DNA hybridization, by both phase contrast AFM and STM, by simply using label-free target DNA, even though no significant changes in the topography of the surface occur after hybridization.

RESULTS AND DISCUSSION

The TS technique was firstly proposed by Hegner et al.[27] for the preparation of Au(111) surfaces having $R_{rms}$ (root mean square roughness) as small as 2-5 Å over several $\mu m^2$, and it is now a widespread method used to grow very smooth inorganic metal films made of Au, Pt or Ag,[28,29] particularly useful as substrates for self assembled monolayers (SAMs), even in ultra high vacuum as recently proposed.[30] The technique is based on the physical vapour deposition (PVD) of a metal film on a freshly cleaved mica surface. The metal layer is then glued onto a silicon wafer and the mica is finally stripped exposing to the ambient the atomic layers which were firstly deposited on the mica surface. The surface of such a film is clean and almost as flat as that of the mica itself. This same procedure was applied to prepare very smooth multi-metal substrates formed by a grid of gold squares with 60 μm sides embedded in a silver background. The squares can be clearly distinguished by an optical microscope, which is of conventional use in most SPM microscopes. The grid also carries an alphabetic identification pattern that allows to identify every single Au square. The process we developed to grow TS multi-metal pattern is composed of four steps as depicted in Fig. 1:

   A. deposition of a gold micro-pattern on a masked mica cleaved surface
   B. deposition of silver over the whole gold pattern (after mask removal)
   C. gluing a silicon wafer on top of the deposited film
   D. mica peeling to uncover the clean surface.

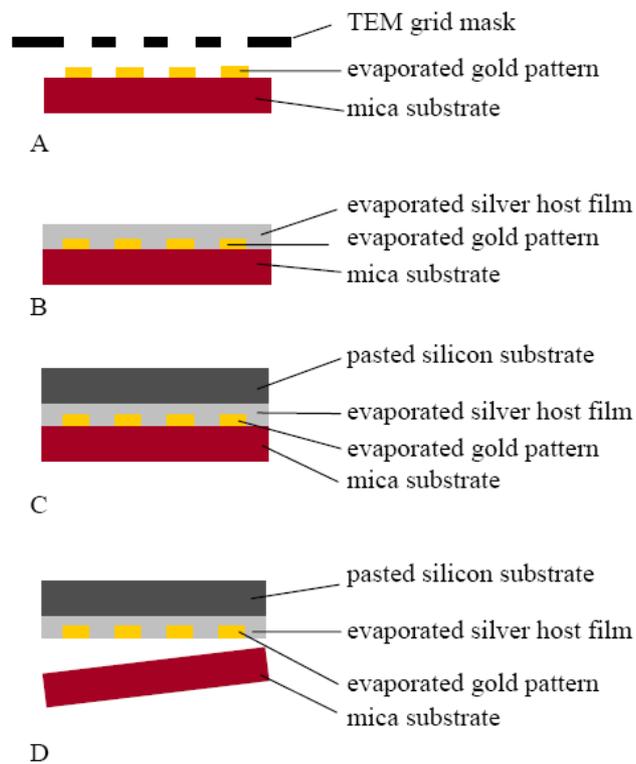

**Figure 1.** Scheme of the fabrication process of the Au-Ag template stripped pattern: PVD evaporation of the gold pattern over the mica substrate (A), coating of the whole substrate by PVD evaporation of a thick film of silver (B), attachment of the silicon wafer on the silver surface (C), peeling of the mica substrate (D).

The metal evaporation was performed in vacuum (pressure $< 2 \cdot 10^{-3}$ Pa) using a coating system equipped with a film thickness monitor. TEM finder grids (200 mesh) were used as masks in the evaporation process and were mounted in contact with the mica surface. 20 nm thick gold dots were evaporated on hot (400 °C) mica with a slow deposition rate (0,01 nm/s) in order to allow homogeneous coverage of the exposed master mica surface. After mask removal a 250 nm thick silver film was condensed on the whole substrate at room temperature. A silicon wafer was then glued on the silver film using a two component epoxy adhesive which required a final thermal annealing at 140 °C in air for 45 minutes. The mica master was removed just before the beginning of DPN experiments by stripping the mica-Au/Ag layer-silicon "sandwich" after a tetrahydrofuran (THF) (Sigma-Aldrich, St. Louis, MO) bath.

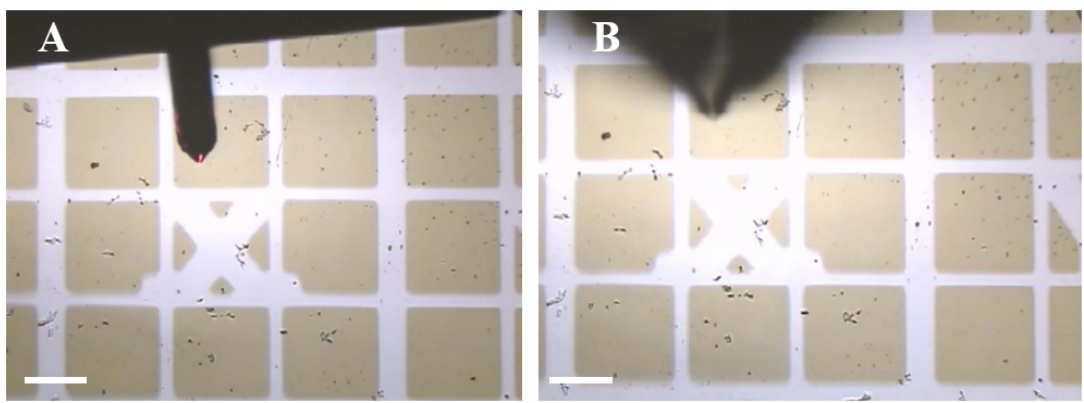

**Figure 2.** Optical image of an AFM cantilever (A) and of a STM electrochemically etched tungsten tip (B) approached over the Au-Ag template stripped pattern (size bar 50 μm). The nine squares associated to the capital letter 'X' are visible in this images.

Silver was chosen to embed the gold pattern because gold and silver can be easily distinguished in optical microscopy for their different colors; mica is also easily pulled off both from silver and gold film with a unique process. The gold dots are incorporated into the silver matrix without the appearance of fractures or bulges on the surface because no internal stress is introduced at the bimetallic interface due to the same cubic close-packed crystal structure with almost identical cell parameters. STM images of gold surface reveal the presence of small, closely packed terraces having size of tens of nanometers, and leading to a smooth homogeneous surface with $R_{rms}$=0.8 ±0.2 nm as measured by contact AFM images taken over 25x25 µm$^2$ scanned areas (not shown) in agreement with measurements on TS gold films.[30,31] The two optical images reported in figure 2 show an AFM cantilever and an electrochemically etched tungsten STM tip approached over the Au/Ag microgrid surface. The precision that one can reach in positioning the probe is of the order of 10 µm for a conventional AFM tip and even less for an electrochemically etched STM tip. Therefore our TS multi-metal patterned substrates allow to easily and precisely approach both an AFM and a STM tip on the site of interest, then the functionalized zone can be simply retrieved by a wide scan around the approached area.

DPN was used on the TS multi-metal patterned substrates for the generation of an oligonucleotide array according to an indirect method.[25] The process could be divided in four main steps:

(1) 1,16-mercaptohexadecanoic acid (MHA) pattern fabrication by DPN;

(2) surface passivation of unpatterned zones with 1-octadecanthiol (ODT);

(3) surface deposition of a 23 bases long amino modified oligonucleotide, and the following generation of the oligonucleotide microarray;

(4) hybridization with the complementary non-functionalized (label-free) oligonucleotide.

In the first step a 3x3 lattice of MHA dots (radius of 2.5 µm) was deposited on the gold surface by dip pen nanolithography using a Nscriptor® DPNWriter equipped with the Nanoink® InkCad lithographic software.

After the generation of the MHA pattern (step 1) the surrounding area was passivated by ODT molecules to limit any possible unspecific binding of the oligonucleotide (step 2).

In step 3 an amino modified oligonucleotide was immobilized onto the MHA spot by the formation of an amide bond; in the final step (step 4) the oligonucleotide was hybridized with a complementary oligonucleotide.

After each step the sample was initially imaged with the same instrument using a clean tip and then moved to a Thermomicroscope Autoprobe CP-research II (Veeco) in order to perform a parallel AFM and STM characterization on a different instrument. Contact mode AFM measurements of samples at step 1 and 3 acquired with the CP instrument are reported in Figure 3. The height of the MHA spots measured by line analysis from topographic images (see fig. 3) is about 1 nm and is comparable with the height of a well formed SAM of MHA.[32] A lower topographic contrast (not shown) is observed in the substrate after the passivation (step 2) since ODT and MHA molecules have a comparable chain length. Moreover after ODT passivation (step 2) the $R_{rms}$ of the surface increases from 0.8 nm to about 2 nm, likely for the presence of some residual impurity of the passivation process, making measurements much more difficult, even though spots can still be identified in phase contrast images (not shown).

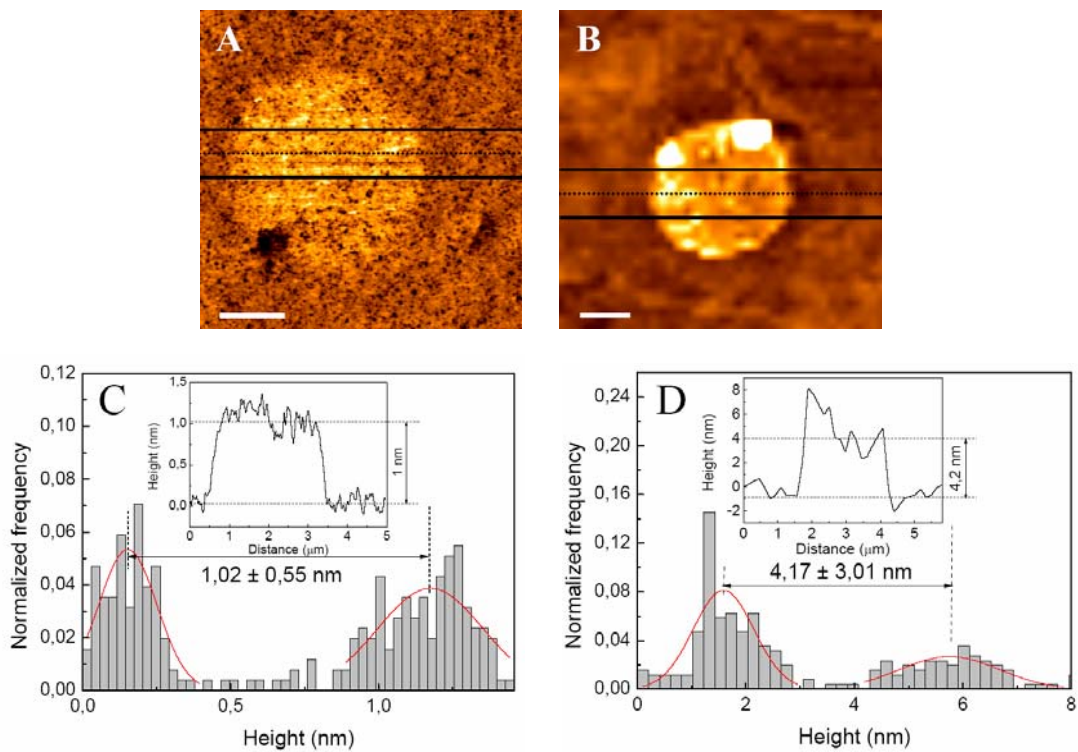

**Figure 3.** Contact-AFM images of a MHA dot (A) and of a oligonucleotide dot (B). Scale bars represent 1 μm in both images. Distributions of the height calculated from the mean height line profiles taken on the MHA dot (C) and on the oligonucleotide dot (D). The mean height line profiles are calculated averaging the line profiles included between the two parallel bold lines reported in the AFM images. The mean line profiles are reported in the insets in (C) and (D).

The spots height after the immobilization of the oligonucleotide (step 3) is about 4 nm. This value is consistent with the presence of a 23-mer oligonucleotide which is about 8 nm long if completely chain extended. Many works in literature have shown that when the oligonucleotide is linked to the surface it is free to bend and so its height is reduced.[33,34] Our height measurements are consistent with a monolayer of 23-mer oligonucleotide molecules inclined with respect to the substrate plane.[20] No significative difference in the topographic image (not shown) and in the height of the spots can be observed after the hybridization (step 4). In spite of the hybridization reaction and the resulting stiffening of the oligonucleotide chain, the spots height does not increase because of the small chain length.

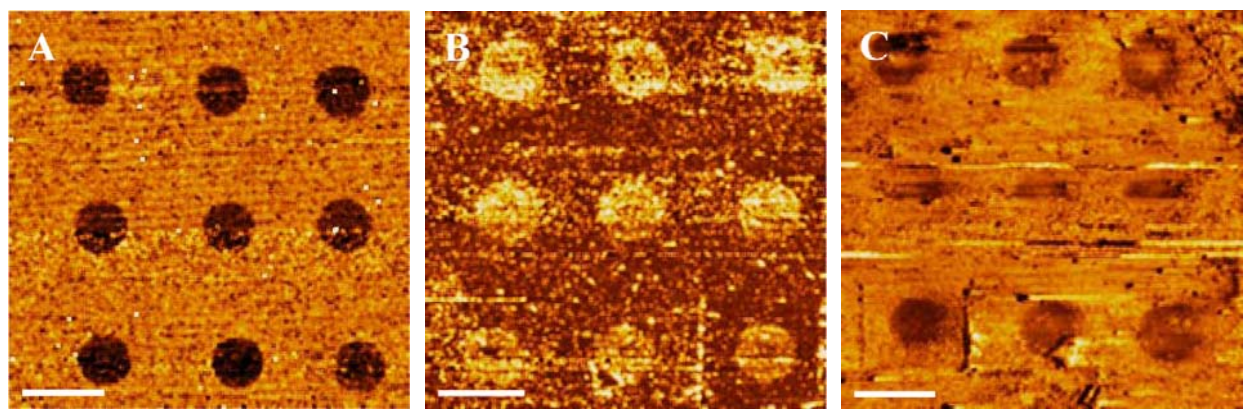

**Figure 4.** Phase contrast images of the DPN generated pattern after writing MHA dots (A), immobilization of amino modified oligonucleotide (B) and hybridization with un-modified complementary oligonucleotide (C). Scale bars represent 5 μm in all images.

Nevertheless, interesting information about occurrence of hybridization can be gained from phase contrast images. AFM phase contrast images of steps (1), (3) and (4) acquired with the Nscriptor immediately after the functionalizations using a clean tip are reported in Figure 4. An inversion of the phase contrast can be observed in the spots zone after the hybridization process (from step 3 to 4) both when measuring with the same instrument used for DPN writing (Nscriptor) and after moving the sample 'ex-situ' to perform AFM/STM measurements (with the Thermomicroscope), as shown in Fig 4 B,C and in Fig 5 A,B. The oligonucleotide appears lighter than the surrounding area, whereas the hybridized oligonucleotide appears darker. This contrast inversion may be related with the modification in the mechanical properties[35] of the oligonucleotide after its hybridization (in particular its rigidity) and could be used as a diagnostic hybridization signal in a future label-free oligonucleotide nanoarray. This remarkable result demonstrates that this detection method does not require any type of modification of the probe DNA molecule.

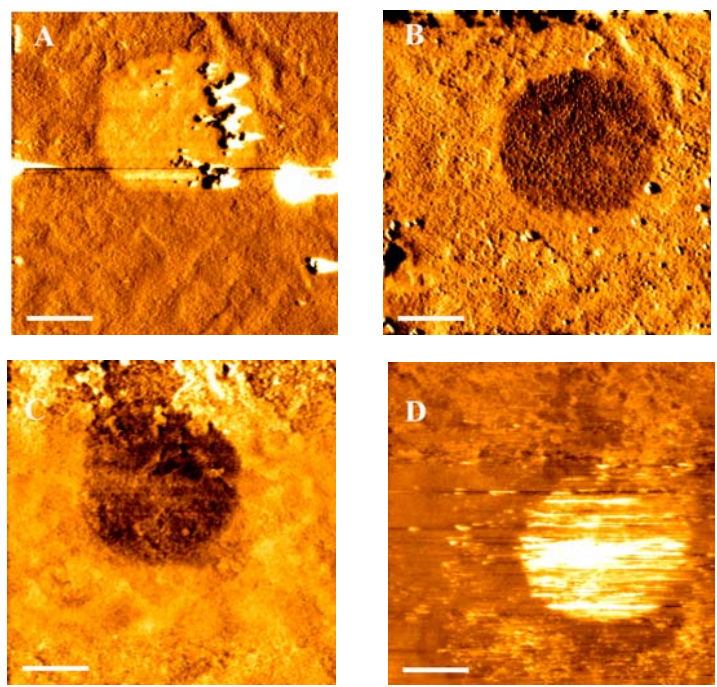

**Figure 5.** Phase contrast images (A, B) and STM topographic images (C, D) of a single amino modified oligonucleotide (A,C) and hybridized (B, D) dot. Scale bars represent 1 μm in all images.

Preliminary STM measurements on the oligonucleotide sample and after hybridization (step 3 and 4) are reported in fig. 5 C,D. Constant current STM images were performed in air using a home made electrochemically etched tungsten tip. As a first indication we observed a different contrast in the topography signal before and after oligonulceotide hybridization and under the same measurement conditions (1.35 nA set point tunneling current and 1.2 V bias voltage). To date, all STM measurements reported in the literature about single or double stranded DNA molecules packed and grafted on a surface were actually acquired on extended self assembled monolayers (SAMs)[36] or alternatively on isolated DNA deposited on atomically flat surfaces,[37] in order to investigate DNA morphology at the molecular scale and electronic properties. Our results open the possibility to perform not only high resolution STM topography but also studies of the electronic properties of DPN-generated patterns, overcoming the difficulties of retrieving the zone of interest; moreover, STM, like phase contrast AFM, is sensitive to oligonucleotide hybridization, and is thus an ideal candidate for sensitive probing of DNA micro- and nanoarrays, along with other proposed detection methods (e.g. Kelvin force microscopy).[14]

CONCLUSIONS

In this work, a novel multi-metal patterned TS substrate was developed and employed as a substrate for scanning probe microscopy investigation of DNA arrays produced by DPN. The optical recognizable micrometric pattern allowed to easily and fast retrieve the DPN functionalized region while the high smoothness of the gold surface over a large scale (0.8 nm ± 0.2 nm over 25x25 μm$^2$) permitted to be sensitive to the local hybridization of the oligonucleotide array by means of different SPM characterizations.

In particular phase contrast AFM and costant current STM images showed optimum sensitivity to hybridization of non-functionalized DNA strands. Our substrate appears thus attractive for the development of DPN-generated label-free DNA nanoarrays with local (high resolution) hybridization detection fulfilled by AFM or tunneling current measurements, and in principle by other scanning microscopy or spectroscopy probes.

MATERIALS AND METHODS

SUBSTRATE PREPARATION

Copper TEM finder grids (200 mesh, purchased from Agar Scientific) used as masks in the evaporation process were mounted in contact with the freshly cleaved surface of a mica plate and a Mo filament heater was fixed on the back of the same mica.

The metal evaporation was executed using an E306A EDWARDS coating system equipped with a quartz microbalance to monitor the film thickness. Before the evaporation of gold, the mica substrate was stored in the evaporation chamber at pressure lower than $2*10^{-3}$ Pa, annealed to 400 °C (temperature monitored by a thermocouple) and kept at this temperature for about one hour to allow the surface degassing.

Silicon wafers were ultrasonically cleaned in iso-propanol for 5 minutes and dried under a high-purity nitrogen stream before being glued to the silver surface.

DNA MICROARRAYS GENERATION BY DIP PEN NANOLITHOGRAPHY

DPN generated MHA arrays were written inside a gold square of the Au-Ag microgrid using a Nscriptor® DPNWriter equipped with the Nanoink® InkCad lithographic software. An AFM tip (Nanoink®, USA) with a spring constant of 0.041 N/m was coated by dipping first in a 5mM MHA acetonitrile solution. The coated tip was then dried with a gentle flux of argon and immediately used for writing ion (Sigma-Aldrich, St. Louis, MO) for 15-20 seconds, and then in deionized water for other 15-20 seconds.

The passivation was obtained by dipping the substrate in a 1mM solution of ODT (Sigma-Aldrich, St. Louis, MO) in ethanol (Sigma-Aldrich, St. Louis, MO) for 5 minutes and then rinsed in ethanol.

In order to attach the 23-base long probe amino modified oligonucleotide (Operon Biotechnologies, Germany) to the DPN generated MHA spots, the surface carboxyl moieties were activated by dipping in an aqueous buffer solution of 1-ethyl-3-(3-dimethylaminopropyl) carbodiimide hydrochloride (EDAC; 10 mg mL$^{-1}$ in 0.1M morpholine/ethanesulfonic acid (MES), pH 4.5) (Sigma-Aldrich, St. Louis, MO) for 15 minutes followed by rinsing with 0.1M sodium borate/boric acid buffer (Sigma-Aldrich, St. Louis, MO), pH 9.5. A droplet (10 μL) of the 25 μM amino modified oligonucleotide solution in borate buffer was deposited on the patterned substrate and let react for 12 hours and then rinsed with phosphate buffered saline (PBS; 0.3M NaCl, 0.01M phosphate, pH 7) (Sigma-Aldrich, St. Louis, MO) and deionized water.

Finally the hybridization was executed placing a complementary target oligonucleotide solution (0.3 μM in PBS buffer) on top of the substrate for 2.5 hours. The substrate was then rinsed with PBS buffer and water and let dry in air.

CHARACTERIZATION (Type of AFM tips, STM tips)

Atomic force microscopy measurements were performed with two different instruments: the Nscriptor used for the DPN functionalization and an Autoprobe CP Thermomicroscope (Veeco). Phase contrast images were realized using silicon probes with nominal force constant and resonance frequency of about 40 N/m and 320 kHz respectively. The contact AFM topography images were acquired with silicon probes characterized by a nominal force constant of 0.2 N/m.

The scanning tunnelling microscopy images were exclusively executed with the Autoprobe CP microscope in ambient conditions. The employed STM tips were home made electrochemically etched tungsten filaments.

ACKNOWLEDGMENT. The authors acknowledge support of this work by Italian Institute of Technology (project Nanobiotechnology "Molecular Imaging"). A.B. acknowledges Ph.D. funding by Italian Institute of Technology.

REFERENCES.

(1) Schena, M.; Shalon, D.; Davis, R. W.; Brown, P. O. *Science* **1995,** 270, (5235), 467-470.

(2) DeRisi, J. L.; Iyer, V. R.; Brown, P. O. *Science* **1997,** 278, (5338), 680-686.

(3) van Hal, N. L. W.; Vorst, O.; van Houwelingen, A.; Kok, E. J.; Peijnenburg, A.; Aharoni, A.; van Tunen, A. J.; Keijer, J. *Journal of Biotechnology* **2000,** 78, (3), 271-280.

(4) DeRisi, J.; Penland, L.; Brown, P. O.; Bittner, M. L.; Meltzer, P. S.; Ray, M.; Chen, Y. D.; Su, Y. A.; Trent, J. M. *Nature Genetics* **1996,** 14, (4), 457-460.

(5) Heller, R. A.; Schena, M.; Chai, A.; Shalon, D.; Bedilion, T.; Gilmore, J.; Woolley, D. E.; Davis, R. W. *Proceedings of the National Academy of Sciences of the United States of America* **1997,** 94, (6), 2150-2155.

(6) Lyubchenko, Y. L. *Cell Biochemistry and Biophysics* **2004**, 41, (1), 75-98.

(7) Alessandrini, A.; Facci, P. *Measurement Science and Technology* **2005**, 16, (6), R65-R92.

(8) Davis, J.J.; Morgan, D.A.; Wrathmell, C.L.; Axford, D.N.; Zhao, J.; Wang, N. *Journal of Materials Chemistry* **2005**, 15, (22), 2160-2174.

(9) Losic, D.; Martin, L.L.; Mechler, A.; Aguilar, M.I.; Small, D.H. *Journal of Structural Biology* **2006**, 155, (1), 104-110.

(10) Shapir, E.; Cohen, H.; Calzolari, A.; Cavazzoni, C.; Ryndyk, D.A.; Cuniberti, G.; Kotlyar, A.; Di Felice, R.; Porath, D. *Nature Materials* **2008**, 7, (1), 68-74.

(11) Ginger, D. S.; Zhang, H.; Mirkin, C. A. *Angew. Chem. Int. Ed.* **2004,** 43, (1), 30-45.

(12) Rosi, N. L.; Mirkin, C. A. *Chemical Reviews* **2005,** 105, (4), 1547-1562.

(13) Lander, E. S. *Nature Genetics* **1999,** 21, 3-4.

(14) Sinensky, A.K.; Belcher, A.M. *Nature Nanotechnology* **2007**, 2, (10), 653-659;

(15) Zhou, D.J.; Sinniah, K.; Abell, C.; Rayment, T. *Angewandte Chemie – International Edition* **2003**, 42, (40), 4934-4937.

(16) Piner, R. D.; Zhu, J.; Xu, F.; Hong, S. H.; Mirkin, C. A. *Science* **1999,** 283, (5402), 661-663.

(17) Hong, S. H.; Zhu, J.; Mirkin, C. A. *Science* **1999,** 286, (5439), 523-525.

(18) Salaita, K.; Wang, Y.H.; Mirkin, C. A. *Nature Nanotechnology* **2007**, 2, (3), 145-155.

(19) Salaita, K.; Wang, Y.; Fragala, J.; Vega, R.A.; Liu, C.; Mirkin, C.A. *Angewandte Chemie International Edition* **2006**, 45, (43), 7220-7223.

(20) Rospigliosi, A.; Ehlich, R.; Hoerber, H.; Middelberg, A.; Moggridge, G. *Langmuir* **2007**, 23, (15), 8264-8271.

(21) Kim, J.M.; Ohtani, T.; Sugiyama, S.; Hirose, T.; Muramatsu, H. *Analytical Chemistry* **2001**, 73, (24), 5984-5991.

(22) Hartschuch, A.; Qian, H.; Meixner, A.J.; Anderson, N.; Novotny, L. *Surface and Interface Analysis* **2006**, 38, (11), 1472-1480.

(23) Vettiger, P.; Despont, M.; Drechsler, U.; During, U.; Haberle, W.; Lutwyche, M.I.; Rothuizen, H.E.; Stutz, R.; Widmer, R.; Binning, G.K. *IBM Journal of Research and Development* **2000**, 44, (3), 323-340.

(24) Yu, X.; Xu, D.; Cheng, Q. *Proteomics* **2006**, 6, 5493-5503.

(25) Demers, L. M.; Park, S. J.; Taton, T. A.; Li, Z.; Mirkin, C. A. *Angew. Chem. Int. Ed.* **2001,** 40, (16), 3071-3073.

(26) Demers, L. M.; Ginger, D. S.; Park, S. J.; Li, Z.; Chung, S. W.; Mirkin, C. A. *Science* **2002,** 296, (5574), 1836-1838.

(27) Hegner, M.; Wagner, P.; Semenza, G. *Surface Science* **1993**, 291, 39-46.

(28) Diebel, J.; Lowe, H.; Samori, P.; Rabe, J.P. *Applied Physics A- Materials Science & Processing* **2001**, 73, (3), 273-279

(29) Blackstock, J.J.; Li, Z.Y.; Freeman, M.R.; Stewart, D.R. *Surface Science* **2003**, 546, (2-3), 87-96.

(30) Lee, S.; Bae, S-S.; Medeiros-Ribeiro, G.; Blackstock, J.J.; Kim, S.; Stewart, D.R.; Ragan, R. *Langmuir* **2008**, 24, (12), 5984-5987.

(31) Wagner, P.; Hegner, M.; Guntherodt, H.J.; Semenza, G. *Langmuir* **1995**, 11, (10), 3867-3875.

(32) Rozhok, S.; Shen, C. K. F.; Littler, P. L. H.; Fan, Z. F.; Liu, C.; Mirkin, C. A.; Holz, R. C. *Small* **2005,** 1, (4), 445-451.

(33) Zhou, D.J.; Sinniah, K.; Albell, C.; Rayment, T. *Langmuir* **2002**, 18, (22), 8278-8281.

(34) Legay G.; Finot, E.; Meunier-Prest, R.; Cherkaoui-Malki, M.; Latruffe, N.; Dereux, A. *Biosensnors and Bioelectronics* **2005**, 21, (4), 627-636.

(35) Bustamante, C.; Smith, S.B.; Liphardt, J.; Smith, D. *Current Opinion in Structural Biology* **2000**, 10, (3), 279-285.

(36) Patole, S.N.;. Pike, A.R; Connolly, B.A.; Horrocks, B.R.; Houlton, A. *Langmuir* **2003**, 19, (13), 5457-5436.

(37) Shapir, E.; Cohen, H.; Borovok, N.; Kotlyar, A.B.; Porath, D. *Journal of Physical Chemistry B*, **2006**, 110, (9), 4430-4433.